\documentclass[fleqn,11pt,double]{wlscirep}
\usepackage[utf8]{inputenc}
\usepackage{graphicx}
\usepackage{cite}
\usepackage{amsmath,amsbsy}
\graphicspath{{images/}}
\usepackage{mathtools}
\usepackage{physics}
\usepackage{algorithm}
\usepackage[noend]{algpseudocode}
\usepackage{caption}
\usepackage{placeins}
\usepackage{todonotes}
\usepackage{subcaption}
\usepackage{newtxtext,newtxmath}
\usepackage[T1]{fontenc}
\usepackage{bm}
\usepackage{multirow}
\newtheorem{lemma}{Lemma}

\title{Efficient Noise Mitigation Technique for Quantum Computing}
\author[1]{Ali Shaib}
\author[1]{Mohamad H. Naim}
\author[2,*]{Mohammed E. Fouda}
\author[1]{Rouwaida Kanj}
\author[2]{Fadi Kurdahi}
\affil[1]{Electrical and Computer Engineering Dept., American University of Beirut, Lebanon, 1107 202.}
\affil[2]{Center for Embedded \& Cyber-physical Systems, University of California-Irvine, Irvine, CA, USA 92697-2625}

\affil[*]{foudam@uci.edu}

\begin{abstract}
Quantum computers have enabled solving problems beyond the current computers' capabilities. However, this requires handling noise arising from unwanted interactions in these systems. Several protocols have been proposed to address efficient and accurate quantum noise profiling and mitigation. In this work, we propose a novel protocol that efficiently estimates the average output of a noisy quantum device to be used for quantum noise mitigation. The multi-qubit system average behavior is approximated as a special form of a Pauli Channel where Clifford gates are used to estimate the average output for circuits of different depths. The characterized Pauli channel error rates, and state preparation and measurement errors are then used to construct the outputs for different depths thereby eliminating the need for large simulations and enabling efficient mitigation. We demonstrate the efficiency of the proposed protocol on four IBM Q 5-qubit quantum devices. Our method demonstrates improved accuracy with efficient noise characterization. We report up to 88\% and 69\% improvement for the proposed approach compared to the unmitigated, and pure measurement error mitigation approaches, respectively.   
\end{abstract}
\begin{document}

\flushbottom
\maketitle
\section*{Introduction}
Building large-scale quantum computers is still a challenging task due to a plethora of engineering obstacles \cite{eng}. One prominent challenge is the intrinsic noise. In fact, implementing scalable and reliable quantum computers requires implementing quantum gates with sufficiently low error rates. There has been substantial progress in characterizing noise in a quantum system \cite{noise3,noise2,noise1} and in building error correcting schemes that can detect and correct certain types of errors \cite{qec1,qec2,qec3}.\\
Numerous protocols have been constructed to characterize the noise in quantum devices. Many of these protocols fail in achieving one of the following desirables: scalability to large-scale quantum computers and efficient characterization of the noise. Quantum Process Tomography \cite{QPT} is a protocol that can give a complete description of the dynamics of a quantum black box, however, it’s not scalable to large-scale quantum systems. Randomized Benchmarking (RB) is another protocol that’s typically used to estimate the error rate of some set of quantum gates \cite{ ScalableNoise, RBQG}. Although RB is a scalable protocol in principle, it can only measure a single error rate that’s used to approximate the average gate infidelity thus providing an incomplete description of noise. Various other protocols based on RB protocol are able to characterize the correlations of noise between the different qubits, however, these protocols lack scalability \cite{ScalableNoise,SimultaneousRB,ThreeRB}.
Quantum Error Mitigation \cite{QEM} (QEM) is a recently emerging field that aims to improve the accuracy of near-term quantum computational tasks. Whereas Quantum Error Correction (QEC) \cite{AQEC1,AQEC2} necessitates additional qubits to encode a quantum state in a multiqubit entangled state, QEM does not demand any additional quantum resources. It is considered an excellent alternative for enhancing the performance of Noisy Intermediate-Scale Quantum (NISQ) computing \cite{nisq}. QEM protocols include zero-noise Richardson extrapolation of results from a collection of experiments of varying noise \cite{Extrapolation}, probabilistic error cancellation through sampling from a set of quantum circuits generated from a target circuit mixed with Pauli gates from an optimized distribution 
\cite{pem,ProbabilisticMitigation}, and exploiting state-dependent bias through invert-and-measure techniques to map the predicted state to the strongest state \cite{bias}.
Measurement Error Mitigation (MEM) is another QEM protocol that models the noise in a quantum circuit as a measurement noise matrix $\bm{E}_{meas}$ applied to the ideal output of the circuit. The columns of $\bm{E}_{meas}$ are the probability distributions obtained through preparing and immediately measuring all possible $2^n$ basis input states \cite{QiskitTextbook}.\\
Recently, the authors in \cite{Nature} developed a protocol based on the RB that relies on the concept of a Gibbs Random Field (GRF) to completely and efficiently estimate the error rates of the Pauli Channel and detect correlated errors between the qubits in a quantum computer. Their effort paves the way to enable quantum error correction and/or mitigation schemes. Herein, we refer to their efficient learning protocol as the \{EL protocol\}. 
In this paper, we build upon the EL protocol and decompose the average noise of a quantum circuit of specific depth into State Preparation and Measurement (SPAM) error and average gate error.
We propose a linear algebraic based protocol and proof to efficiently construct and model the average behavior of noise in a quantum system for any desired circuit depth without having to run a large number of quantum circuits on the quantum computer or simulator. We then rely on this model to mitigate the noisy output of the quantum device. For an n-qubit quantum system, the average behavior of the noise can be well approximated as a special form of a Pauli Channel \cite{Knill2005,Wallman_2016,Ware_2021}. 
A Pauli channel $\varepsilon$ acts on a qubit state $\boldsymbol{\rho}$ to produce
\begin{equation}
\varepsilon(\bm{\rho})=\sum_{i}p_i\bm{P}_{i}\bm{\rho}\bm{P}_{i}
\label{equation 1}
\end{equation}
where $p_i$ is an error rate associated with the Pauli operator $\bm{P}_{i}$. The $p_i$'s form a probability distribution $(\sum_{i}p_{i}=1)$, and are related to the eigenvalues, $\boldsymbol{\lambda}$, of the Pauli Channel defined as 
\begin{equation}
\lambda_i=2^{-n}Tr(\bm{P}_{i}\varepsilon(\bm{P}_{i}))
\label{Equation 2}
\end{equation}
Thus, when a state $\bm{\rho}$ is subjected to the noisy
channel $\varepsilon$, $p_i$ describes the probability of a multiqubit Pauli error $\bm{P}_{i}$ affecting the system, while $\lambda_i$ describes how faithfully a given multispin Pauli operator is transmitted. $\bm{p}$ and $\bm{\lambda}$ are related by Walsh-Hadamard transform where

\begin{equation}
\bm{\lambda}=\bm{Wp}
\label{Equation 3}
\end{equation}

While RB only estimates the average value of all $\lambda_i$ of the Pauli Channel, the EL protocol estimates the individual $\lambda_i$. A complete characterization of the Pauli channel requires learning more than the eigenvalues or error rates associated with single-qubit Pauli operators such as $\bm{\sigma}_{z}^{(1)}$ or  $\bm{\sigma}_{x}^{(3)}$; it requires learning all of the noise correlations in the system, that is, also learning the eigenvalues and error rates associated with multiqubit Pauli operators such as $\bm{\sigma}_{z}^{(1)} \otimes \mathbf{1}^{(2)} \otimes \bm{\sigma}_{x}^{(3)}$ and how they vary compared to the ones obtained under independent local noise. Estimating these correlations is essential for performing optimal QEC and/or QEM. However, these  correlations increase exponentially as the number of qubits increases, so having an efficient noise characterization protocol is crucial to direct the error mitigation efforts to capture the critical noise correlations.\\
Our method relies on the error rates vector $\bm{p}$ of the Pauli-Channel to decompose the average behavior of noise for circuits of depth $m$ into two noise components: a SPAM error matrix denoted by the matrix $\bm{N}$ and a depth dependent component comprising an average gate error matrix denoted by the matrix $\bm{M}$. We evaluate our model for the average noise by predicting the average probability distribution for circuits of depth $m$ and computing the distance between this predicted distribution and the empirically obtained one. Finally, we use our proposed decomposition to mitigate noisy outputs of random circuits and compare our mitigation protocol with the MEM protocol \cite{QiskitTextbook}. We applied our noise characterization and mitigation protocols on the following IBM Q 5-qubit quantum computers: Manila, Lima, and Belem\cite{IBM}.
\section*{Results}
\subsection*{Proposed Protocol Theory}
The ideal output probability distribution of an $n$-qubit quantum circuit with depth $m$ is perturbed by the SPAM and the average gate errors. Our aim is to construct a comprehensive linear algebraic model that takes into account both these errors for an arbitrary depth $m$. Matrix algebra can then be employed to mitigate the noise as follows: 
\begin{equation}
    \bm{C}_{ideal}= \bm{Q}_{m}^{-1}\bm{C}_{noisy}
\end{equation}
where $\bm{Q}_{m}$ is the characterized noise matrix for circuits of depth $m$, $\bm{C}_{ideal}$ and $\bm{C}_{noisy}$ are the ideal and noisy outputs of a given circuit of depth $m$, respectively.
The straight-forward approach would be to construct $\bm{Q}_m$ from empirical simulations in a similar fashion to the $\bm{E}_{meas}$ noise matrix that was characterized in the MEM scheme. The columns of $\bm{Q}_{m}$ comprise the emperical average probability distributions for basis input states $\ket{\bm{in}}\in \{ \ket{\bm{0}},\,\ket{\bm{1}},\,\dots,\,\ket{\bm{2^n-1}}\}$, denoted by $\hat{\bm{q}}(m,\ket{\bm{in}})$, where $\hat{\bm{q}}(m,\ket{\bm{in}})$ are obtained through sampling a number of depth $m$ circuits to incorporate the average gate and SPAM errors.
\begin{equation}
\bm{Q}_m=
\begin{bmatrix}
 \hat{\bm{q}}(m,\ket{\bm{0}}) & \hat{\bm{q}}(m,\ket{\bm{1}}) & \hdots &\hat{\bm{q}}(m,\ket{\bm{2^n-1}})
\end{bmatrix}
\label{eq:Q}
\end{equation}
Building $\bm{Q}_m$, however, through empirical simulations can be expensive especially when the circuit depth is large. Herein, we propose a method for an efficient estimation of $\bm{Q}_m$ where the individual probability distributions $\hat{\bm{q}}(m,\ket{\bm{in}})$ are estimated as follows:
\begin{equation}
    \bm{q}'(m, \ket{\bm{in}}) = \bm{N}_{in}\bm{M}_{in}^m\ket{\bm{in}}
    \label{eq:qhat}
\end{equation}
where $\bm{N}_{in}$ and $\bm{M}_{in}$ are input-specific matrices that represent the SPAM error matrix and average gate error for input $\ket{\bm{in}}$, respectively. Both $\bm{M}_{in}$ and $\bm{N}_{in}$ are extracted empirically using random circuits from a set of small circuit depths $T$ and then used in mitigating the outputs for circuits with higher depths. We first show the construction of $\bm{N}_{0}$ and $\bm{M}_{0}$.\\
The construction of $\bm{N}_{0}$ and $\bm{M}_{0}$ proceeds by estimating the error rates vector $\bm{p}$ associated with the Pauli Channel based on the assumption in Equation \ref{equation 1} for the average behavior of the noisy quantum device at hand using the EL protocol. The protocol proceeds by constructing $K$ random identity circuits of depth $m \in T$ \cite{SimultaneousRB, Nature}. Each circuit is constructed by initializing the qubits to the all-zeros state $\ket{\bm{0}}$ followed by choosing a random sequence $s \in S_{m}$, the set of all length $m$ sequences of one-qubit Clifford gates applied independently on each qubit, followed by an inverse gate for the chosen sequence to ensure an identity circuit. It then estimates the resulting empirical probability distribution $\hat{\bm{q}}(m,\ket{\bm{0}})$ by averaging over all the empirical probability distributions $\hat{\bm{q}}(m,s,\ket{\bm{0}})$ for the constructed random identity circuits of depth $m$, that is, 
\begin{equation}
    \hat{\bm{q}}(m,\ket{\bm{0}})= \frac{1}{K}\sum \hat{\bm{q}}(m,s,\ket{\bm{0}})
    \label{Equation 4}
\end{equation}
$\hat{\bm{q}}(m,\ket{\bm{0}})$ is a vector with $2^n$ entries each corresponding to the possible observed outcome. A Walsh-Hadamard transform is then applied on each $\hat{\bm{q}}(m,\ket{\bm{0}})$ to obtain
\begin{equation}
    \bm{\Lambda}(m)=\bm{W}\hat{\bm{q}}(m,\ket{\bm{0})}
    \label{Equation 5}
\end{equation}
Each parameter $\Lambda_{i}(m)$ in $\bm{\Lambda}(m)$ is fitted to the model 
\begin{equation}
    \Lambda_{i}(m)=A_{i}\lambda_{i}^{m}
    \label{Equation 6}
\end{equation}
where $A_i$ is a constant that absorbs SPAM errors and 
the vector $\bm{\lambda}$ of all fitted parameters $\lambda_i$ is a SPAM-free estimate to the eigenvalues of the Pauli Channel defined in Equation \ref{Equation 2}. Notice that we can rewrite Equation \ref{Equation 6} as
\begin{equation}
    \bm{\Lambda}(m)=\bm{A\lambda}^{m}
    \label{Equation 7}
\end{equation}
where $\bm{A}$ is a diagonal matrix where the diagonal entries are $A_i$ and $\bm{\lambda}^m$ is an element-wise exponentiation of a vector. An inverse Walsh-Hadamard Transform is then applied on $\bm{\lambda}$ to get the error rate vector $\bm{p}$ of the Pauli Channel as
\begin{equation}
    \bm{p}=\bm{W}^{-1}\bm{\lambda}
    \label{Equation 8}
\end{equation}
$\bm{p}$ is then projected onto a probability simplex to ensure $\sum_{i}p_i=1$. Introducing the GRF model by the EL protocol allows the scalability of estimating $\bm{p}$ with the increase in the number of qubits. The GRF model assumes the noise correlations are bounded between a number of neighboring qubits depending on the architecture of the quantum computer at hand. Thus, decreasing the number of noise correlations to be estimated.\\
The final outcome $\bm{p}$ of the EL protocol represents the SPAM-free probability distribution of the average noise in the quantum computer. Each element $p_i \in \bm{p}$ corresponds to the probability of an error of the form $binary(i)$ on an input state $\ket{\bm{0}}$. For example, for a 5-qubit quantum computer, $p_0$ corresponds to the probability of no bit flips on the input state, i.e., error of the form  $IIIII$, $p_1$ to the error of the form $IIIIX$, $p_2$ to the error of the form $IIIXI$, etc…\\
In order to proceed with the proof for our proposed decomposition of Equation (\ref{eq:qhat}) for input state $\ket{\bm{0}}$, we first state the following lemma (the detailed proof of the lemma can be found Section I in the supplementary):
\begin{lemma} 
Let $\bm{\lambda}$ and $\bm{p}$ be the respective eigenvalues and error rates of a Pauli Channel with $n$ qubits, then $\bm{\lambda}^m=\bm{WM}^m\ket{\bm{0}}$
where $\bm{M}$ is a $2^n \times 2^n$ matrix such that $M_{ij}=p_{i\oplus j}$ ($i \oplus j$ is the bitwise exclusive-OR operator).
\label{Lemma}
\end{lemma}
\noindent Using Lemma \ref{Lemma} and Equations \ref{Equation 5} and \ref{Equation 7}, $\hat{q}(m,\ket{0})$ can be estimated as
\begin{equation}
    \bm{q}'(m,\ket{\bm{0}})=\bm{W}^{-1}\bm{AWM}^{m}\ket{\bm{0}}
    \label{Equation 11}
\end{equation}
The transition matrix $\bm{M}=\bm{M}_{0}$ represents the average error per gate while the $\bm{N}=\bm{W}^{-1}\bm{AW}=\bm{N}_{0}$ matrix represents the SPAM errors for an input state $\ket{\bm{0}}$. Notice that the average noise for depth $m$ circuits on an input state $\ket{\bm{0}}$ behaves as a sequence of $m$ average noise gates $\bm{M}_{0}$ followed by SPAM errors $\bm{N}_{0}$.\\
The construction of $\bm{N}_{in}$ and $\bm{M}_{in}$ for input state $\ket{\bm{in}}$ proceeds similar to the procedure of constructing $\bm{N}_{0}$ and $\bm{M}_{0}$, however, a permutation of $\hat{\bm{q}}(m,\ket{\bm{in}})$ is required before applying a Walsh-Hadamard transform to ensure that each element $p_{i}(\ket{\bm{in}})$ in the input-specific error rate vector $\bm{p}(\ket{\bm{in}})$ corresponds to the probability of an error of the form $binary(i)$ on an input state $\ket{\bm{in}}$. This permutation is done by applying an input-specific permutation matrix $\bm{\pi}_{in}$ on $\hat{\bm{q}}(m,\ket{\bm{in}})$ $\forall m$ where $\pi_{in_{ij}}=1$ if $i\oplus j=in$ and $0$ otherwise. 

\subsection*{Experiments}
In this section, we evaluate the accuracy of the model in Equation $\ref{Equation 11}$ in predicting the average probability output, $\hat{\bm{q}}(m,\ket{\bm{0}})$, for identity circuits of higher depths by estimating $\bm{A}_0$ and $\bm{p}(\ket{\bm{0}})$ using only simulations of lower depths identity circuits. Denote by $\bm{q}'(m,\ket{\bm{0}})$ the predicted average probability distribution obtained using Equation \ref{Equation 11}. We select a \textit{training set of depths} $T=\{1,\,2,\,\dots,\,m_{max}\}$ to estimate $\bm{A}_0$ and $\bm{p}$ using the EL protocol followed by the construction of the average gate error matrix $\bm{M}_{0}$ and SPAM error matrix $\bm{N}_{0}$ where $M_{0_{ij}}=p_{i \oplus j}(\ket{\bm{0}})$ and $\bm{N}_{0}=\bm{W}^{-1}\bm{A}_{0}\bm{W}$. A new \textit{testing set of depths} $T'=\{m_{max}+1,\,m_{max}+2,\,\dots,\,100\}$ is then selected where we compute the \textit{Jensen-Shannon Divergence} ($JSD$) between $\hat{\bm{q}}(m',\ket{\bm{0}})$ and $\bm{q}'(m',\ket{\bm{0}})$ $\forall m'\in T'$. The $JSD$ measures the similarity between the two probability distributions \cite{JSD}. The lower the $JSD$, the closer the two distributions are. More information about the $JSD$ can be found in Section II in the supplementary. Figure \ref{VaryingTrainingDepth} presents the computed $JSD$ for different quantum computers while varying $m_{max}$. Figure \ref{TestErrorBars} presents the average and standard deviation for the test $JSD$ values for the different quantum computers. The average test $JSD$ varies between $0.024$ and $0.056$ for the different $m_{max}$ values with lower average $JSD$ values noted for high $m$ for $m_{max}=80$ as indicated in Figure \ref{TestErrorBars}b.
\begin{figure}[h]
\begin{subfigure}[t]{0.32\textwidth}
  \centering
  \includegraphics[width=\columnwidth]{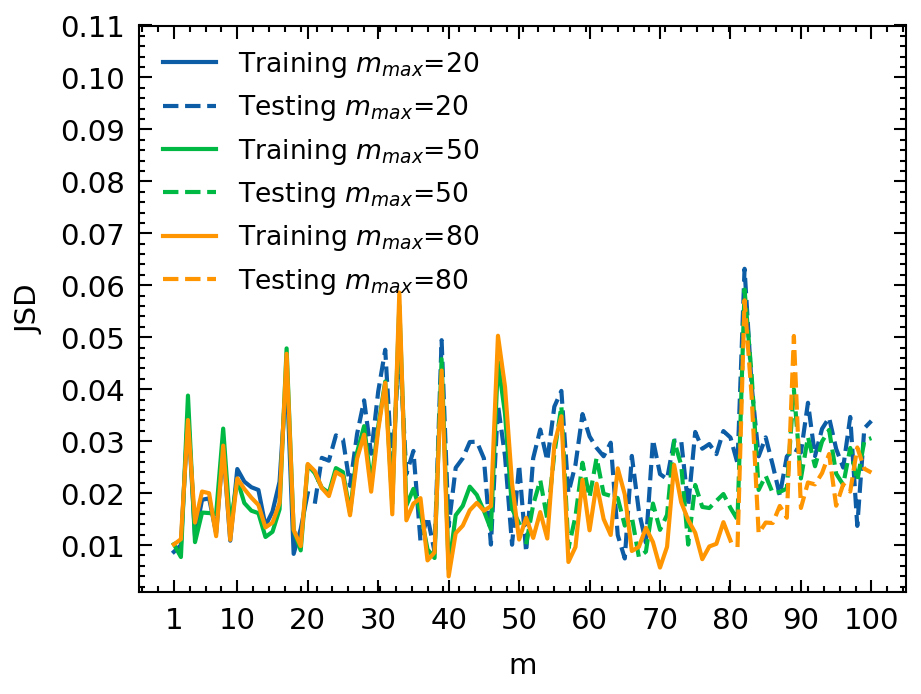}
  \caption{IBM Q Manila}
  \label{Manila}
\end{subfigure}
\begin{subfigure}[t]{0.32\textwidth}
  \centering
  \includegraphics[width=\columnwidth]{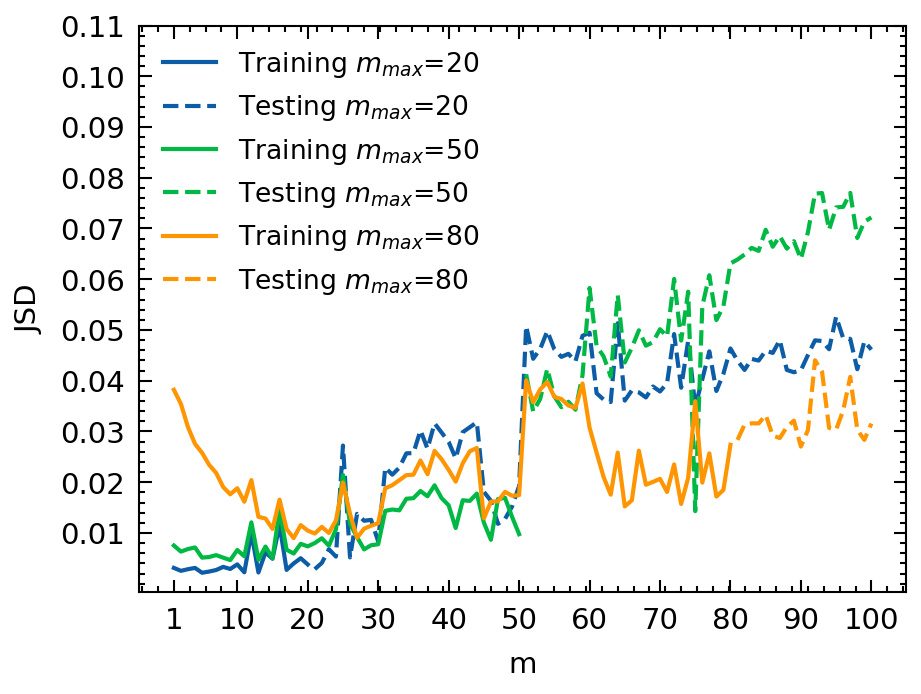}
  \caption{IBM Q Lima}
  \label{Lima}
\end{subfigure}
\begin{subfigure}[t]{0.32\textwidth}
  \centering
  \includegraphics[width=\columnwidth]{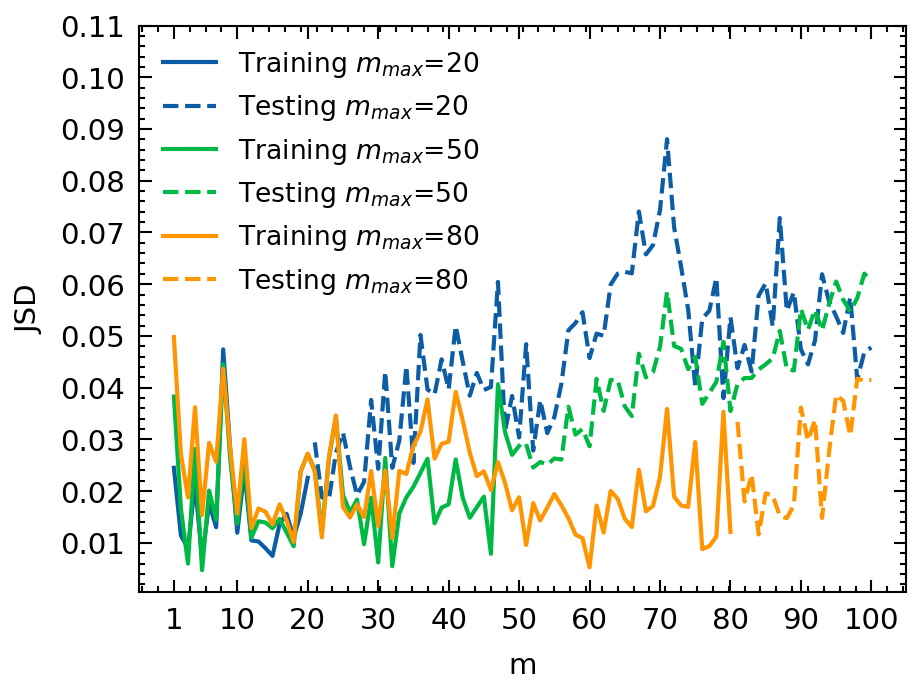}
  \caption{IBM Q Belem}
  \label{Belem}
\end{subfigure}
\caption{$JSD(\hat{\bm{q}}(m,\ket{\bm{0}}),\,\bm{q}'(m,\ket{\bm{0}}))$ for training sets of depths $T$ and testing sets of depths $T'$ with variable maximum training depth $m_{max}\in\{20,\,50,\,80\}$ on different IBM Q 5-qubit quantum computers.}
\label{VaryingTrainingDepth}
\end{figure}
\FloatBarrier
\begin{figure}[h]
    \centering
    \begin{subfigure}[t]{0.48\textwidth}
      \centering
      \includegraphics[width=\columnwidth]{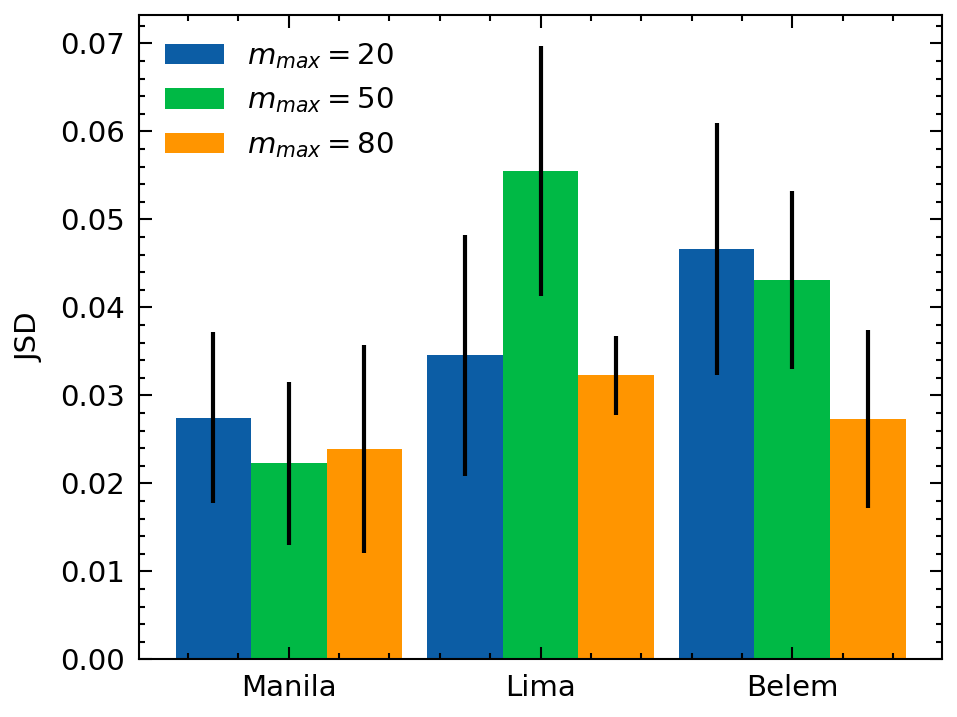}
      \caption{}
    \end{subfigure}
    \begin{subfigure}[t]{0.48\textwidth}
      \centering
      \includegraphics[width=\columnwidth]{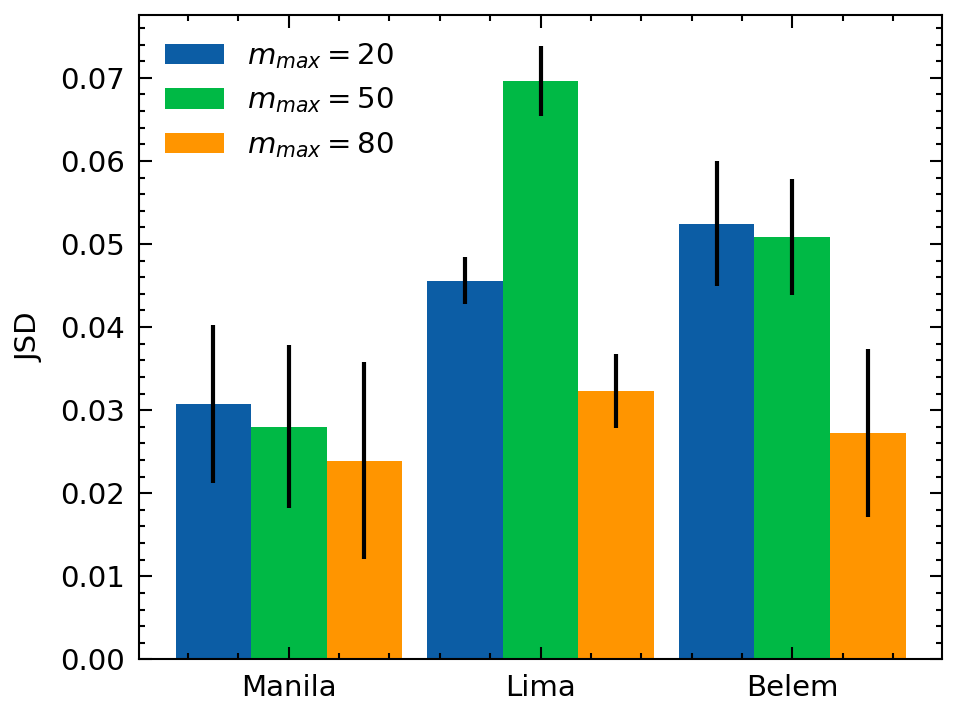}
      \caption{}
    \end{subfigure}
    \caption{The average and standard deviation of $JSD(\hat{\bm{q}}(m,\ket{\bm{0}}),\,\bm{q}'(m,\ket{\bm{0}}))$; (a) over all depths $m \in [m_{max}+1,100]$ and  (b) over depths $m \in [80,100]$ while varying the maximum training depth $m_{max}$ on different IBM Q 5-qubit quantum computers.}
    \label{TestErrorBars}
\end{figure}
\FloatBarrier
\noindent We rely on $\bm{q}'(m, \ket{\bm{in}})$ to construct and evaluate the mitigation power of $\bm{Q}_m$ for different depths. We first  select a \textit{training set of depths} $T=\{1,\,20,\,40,\,60,\,80,\,100\}$ to estimate $\bm{A}_{in}$ and $\bm{p}(\ket{\bm{in}})$ for each input state $\ket{\bm{in}}$ using the EL protocol followed by the construction of $\bm{M}_{in}$ using $\bm{p}(\ket{\bm{in}})$ and $\bm{N}_{in}=\bm{W}^{-1}\bm{A}_{in}\bm{W}$. We then estimate $\hat{\bm{q}}(m,\ket{\bm{in}})$ as $\bm{q}'(m,\ket{\bm{in}})$ for all inputs using Equation \ref{eq:qhat} in order to construct $\bm{Q}_{m}$ using Equation \ref{eq:Q}. We then choose a new \textit{testing set of depths} $T'=\{10,\,30,\,50,\,70,\,90\}$ so that $\bm{Q}_m$ is used in mitigating the outputs for circuits of depth $m \in T'$ where for a given identity circuit of depth $m$ with input $\ket{\bm{in}}$ and sequence $s$ of gates, the mitigated output $\hat{\bm{q}}(m,s,\ket{\bm{in}})_{mit}$ in obtained as 
\begin{equation}
    \hat{\bm{q}}(m,s,\ket{\bm{in}})_{mit}=\bm{Q}_{m}^{-1}\hat{\bm{q}}(m,s,\ket{\bm{in}})
\end{equation}
$\hat{\bm{q}}(m,s,\ket{\bm{in}})_{mit}$ is projected onto a probability simplex to ensure a probability distribution. The $JSD$ between $\hat{\bm{q}}(m,s,\ket{\bm{in}})_{mit}$ and the ideal output $\ket{\bm{in}}$ is computed and then averaged over all input states and all random circuits of depth $m$. We also compare our proposed mitigation protocol using $\bm{Q}_m$ with the MEM scheme (Figure \ref{Mitigation}). We report upto 88\% improvement in the $JSD$ value for the proposed approach compared to the unmitigated approach, and upto 69\% improvement compared to MEM approach. Note that for the results presented here, we rely on the average SPAM free error rate, $\bm{p}_{avg}=\frac{1}{2^n}\sum_{in=0}^{2^n-1}{\bm{p}(\ket{\bm{in}})}$ to construct $\bm{M}_{in}=\bm{M}_{avg}$ for all inputs. We compare the results using $\bm{p}_{avg}$ and $\bm{p}(\ket{\bm{in}})$ in the supplementary Section V. $\bm{N}_{in}$ remains input specific. Further elaborations on the results are presented in supplementary Section VI. 
\begin{figure}[h]
\centering
\begin{subfigure}[h]{0.32\textwidth}
  \centering
  \includegraphics[width=\columnwidth]{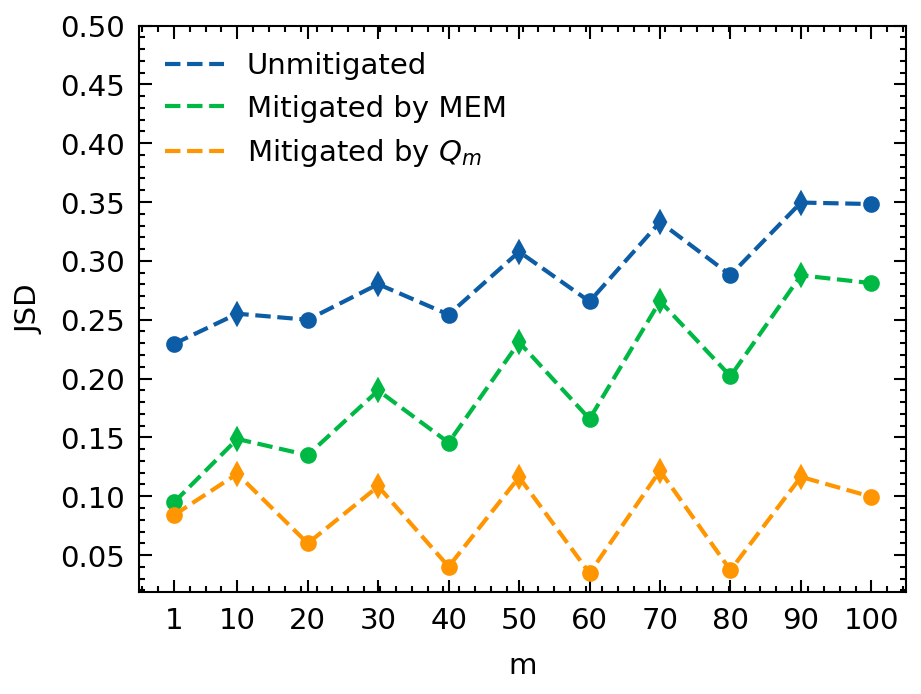}
  \caption{IBM Q Manila}
  \label{LimaMitigation}
\end{subfigure}
\begin{subfigure}[h]{0.32\textwidth}
  \centering
  \includegraphics[width=\columnwidth]{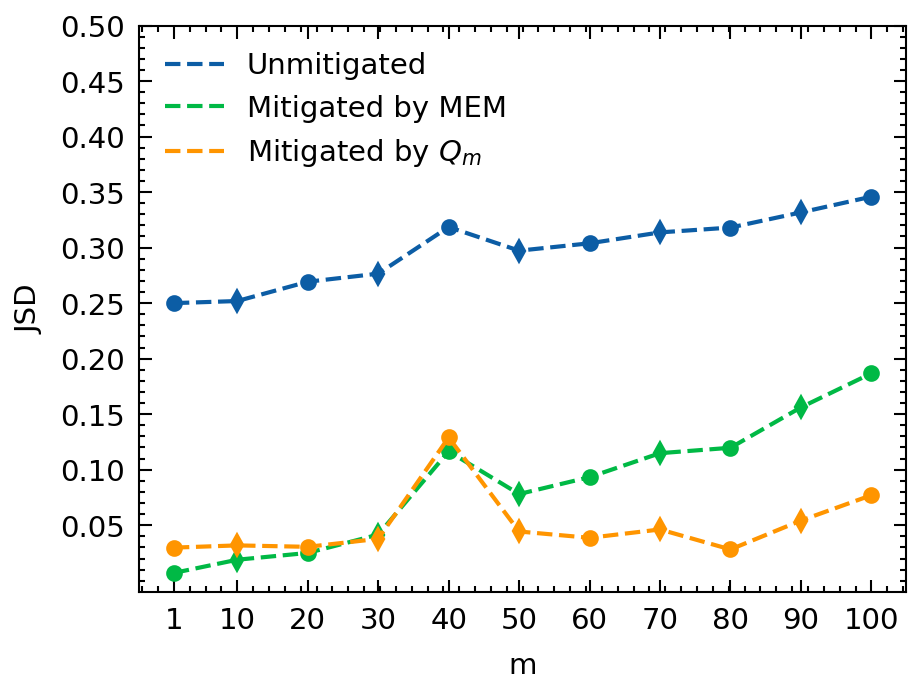}
  \caption{IBM Q Lima}
  \label{AthensMitigation}
\end{subfigure}
\begin{subfigure}[h]{0.32\textwidth}
  \centering
  \includegraphics[width=\columnwidth]{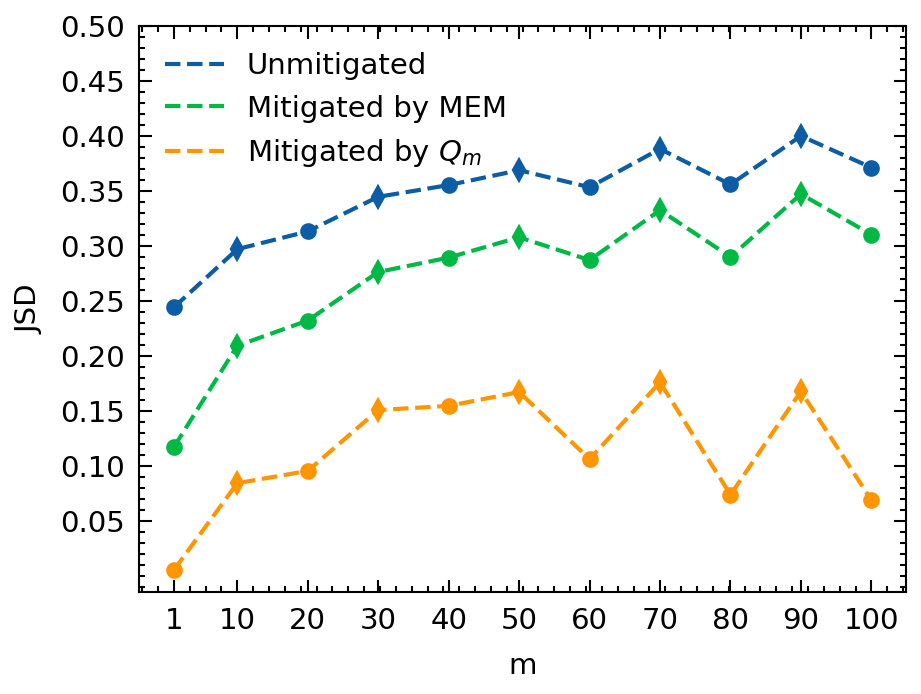}
  \caption{IBM Q Belem}
  \label{belem Mitigation}
\end{subfigure}
\caption{Average $JSD$ between the ideal output $\ket{\bm{in}}$ and each of the unmitigated output $\hat{\bm{q}}(m,s,\ket{\bm{in}})$, mitigated output by the MEM protocol, and mitigated output by our proposed noise model for each depth $m$ on IBM Q 5-qubit quantum computers.}
\label{Mitigation}
\end{figure}
\FloatBarrier
\subsection*{Complexity}
So far in the estimation of $\bm{M}_{in}$ and $\bm{N}_{in}$ for each input state $\ket{\bm{in}}$ using the EL protocol, $K$ random circuits are generated for each depth $1\le m \le m_{max}$ where the EL protocol requires $O(2^{2n})$ for the Walsh-Hadamard transform which can be reduced into $O(n^2)$ using fast Walsh-Hadamard transform. Thus, the overall complexity of the construction of $\bm{M}_{in}$ and $\bm{N}_{in}$ for all input states is $O(m_{max}Kn^22^{n})$. Furthermore,  the GRF model factors the error rates vector into a product of $f\sim O(n)$ factors, depending on the architecture of the quantum computer, where each factor depends on a subset of adjacent qubits of cardinality $N<<n$ (typically $N=4$). Thus, the complexity is reduced further into $O(nm_{max}KN^{2}2^{n})$. The construction of $\bm{Q}_m$ would be based on Equation \ref{eq:qhat} for each input state $\ket{\bm{in}}$ where $\bm{M}_{in}^m$ can be computed efficiently using the Singular Value Decomposition (SVD) of $\bm{M}_{in}$, thus the construction of $\bm{Q}_m$ is $O(2^{3n})$. For the MEM scheme, the construction of $\bm{E}_{meas}$ requires only generating $K$ circuits with no gates for each input state $\ket{\bm{in}}$, thus the complexity is $O(K2^{n})$. For mitigation, both protocols are based on matrix inversion, thus the complexity for mitigation is $O(2^{3n})$.

\section*{Discussion}
The proposed mitigation protocol builds upon the SPAM-free noise characterization protocols for low circuit depths to generate a SPAM-error matrix $\bm{N}_{in}$ and an average gate error matrix $\bm{M}_{in}$ for each input state $\ket{\bm{in}}$. It then constructs a noise mitigation matrix $\bm{Q}_{m}$ for arbitrary circuit depths $m$ where the columns of $\bm{Q}_{m}$ are the estimated average probability distributions $\bm{q'}(m,\ket{\bm{in}})=\bm{N}_{in}\bm{M}_{in}^m$.
The mitigated output $\hat{\bm{q}}(m,\,s)_{mit}$ of a given circuit of depth $m$ with sequence $s$ of gates is obtained by applying ${\bm{Q}_{m}}^{-1}$ on the empirical circuit output $\hat{\bm{q}}(m,\,s)$.



We evaluated the accuracy of our model in estimating the average probability distributions for high depth circuits and evaluated our mitigation protocol on the IBM Q 5-qubit quantum devices: Belem, Lima, and Manila.
For the model accuracy evaluations, for the different $m_{max}$ values, we reported on average a test  $JSD(\hat{\bm{q}}(m,\ket{\bm{0}}),\bm{q}'(m,\ket{\bm{0}}))$ value around 0.022-0.028 for Manilla, 0.03-0.055 for Lima, and 0.028-0.048 for Belem. For $m_{max}=20$, the test JSD values varied between 0.005 and 0.05 for Lima computer, 0.01 and 0.06 for Manila computer, and 0.02 and 0.09 for Belem computer. We note that for $m_{max}=20$ the test spans $m \in [21-100]$. For higher depths $m \in [80-100]$, on average $m_{max}=80$ resulted in better model error than $m_{max}=50$ and $m_{max}=20$. Results for IBM Q Athens are presented in the supplementary.

Finally, we report upto 88\% JSD improvement for the proposed approach compared to the unmitigated approach with significant mitigation improvement compared to MEM at mid to higher depths. Specifically, for $m=90$, we reported 58\%, 66\% and 85\% JSD improvement for the proposed approach compared to the unmitigated on Belem, Manilla, and Lima respectively. This is compared 12\%, 17\% and 51\% respectively for the MEM. On average for all the test depths across the different machines, we report 68.4\% JSD improvement for the proposed versus versus 38.2\% for the MEM improvement compared to the unmitigated approach.
\section*{Methods}
In evaluating the accuracy of the model, we run $K=1000$ random identity circuits with each submission requesting 1024 shots for each depth $m \in \{1,\,2,\,\dots,\,100\}$. In evaluating the mitigation power of $\bm{Q}_m$, we run $K=1000$ random identity circuits for depths $m \in \{1,\,10,\,20,\,30,\,40,\,50,\,60,\,70,\,80,\,90,\,100\}$ for each basis input state $\ket{\bm{in}}$ with each circuit requesting 1024 shots. The constructed circuits contain single qubit Clifford gates only. We run the circuits on the following IBM Q 5-qubit quantum computers: Manila, Lima, and Belem. Theoretical derivations and numerical details essential to the study are presented in the Results section. More details can be found in the Supplementary Information. For the configurations and noise profiles of the IBM quantum machines, please go to IBM Quantum Experience at \url{http://www.research.ibm.com/quantum}.

\subsection*{Data availability}
The data that supports the findings of this study are available from the corresponding author upon reasonable request.

\bibliography{main}


\section*{AUTHOR CONTRIBUTIONS}

A.S., M.F. and R.K. conceived the idea and designed the study. A.S. and M.N. collected the data and run the simulations. A.S. and R.K. prepared the manuscript with input and critical feedback from all authors. All authors analyzed and discussed the results. All authors critically reviewed and developed the final manuscript.

\section*{COMPETING INTERESTS}
The authors declare no competing interests

\newpage

\begin{center}
\LARGE Supplementary Information
\end{center}

This supplementary document supports the discussion in the main text by providing theoretical proof and technical details. Section 1 provides the proof of Lemma 1. Section 2 provides information about the Jensen-Shannon Divergence ($JSD$). Section 3 summarizes our proposed algorithm for the construction of $Q_m$. Section 4 compares the performance of $Q_m$ between using $p(\ket{\bm{in}})$ and $p_{ave}$. Section 5 presents the average $JSD$ and standard deviation for each input state $\ket{\bm{in}}$ over different depths.

\section*{Proof of Lemma 1}

\begin{lemma} 
Let $\boldsymbol{\lambda}$ and $\bm{p}$ be the respective eigenvalues and error rates of a Pauli Channel with $n$ qubits, then 
\begin{equation*}
    \boldsymbol{\lambda}^m=\bm{WM}^m\ket{\bm{0}}
\end{equation*}
where $\bm{W}$ is a Walsh-Hadamard Transform, and $\bm{M}$ is a $2^n \times 2^n$ matrix such that $M_{ij}=p_{i\oplus j}$ ($ \oplus $ is the bitwise exclusive-OR operator).
\label{Lemma}
\end{lemma}


\noindent \textit{Proof of Lemma \ref{Lemma}}:
\noindent For the LHS of Lemma \ref{Lemma}, we have
\begin{equation}
    \boldsymbol{\lambda}=\bm{Wp}
\end{equation}

\noindent Then
\begin{equation}
    \lambda_i=\sum_{k=0}^{2^n-1}(-1)^{ik}p_k
\end{equation}

\noindent And,
\begin{equation}
    \lambda_{i}^{m}=\left(\sum_{k=0}^{2^n-1}(-1)^{ik}p_k\right)^{m}
\end{equation}

\noindent For the RHS of Lemma \ref{Lemma}, using Lemma \ref{Lemma2}, we get
\begin{align}
    \bm{WM}^m\ket{\bm{0}}
    &=\bm{W}(\frac{1}{2^n}\bm{W}\boldsymbol{\Gamma} \bm{W})^m\ket{\bm{0}}
    =\bm{W}(\frac{1}{2^n}\bm{W}\boldsymbol{\Gamma}^m \bm{W})\ket{\bm{0}} \nonumber \\
    &=\boldsymbol{\Gamma}^m \bm{W}\ket{\bm{0}}
    =\boldsymbol{\Gamma}^m \Vec{\mathbf{1}}
\end{align}

\noindent Then
\begin{align}
    (WM^m\ket{\bm{0}})_{i}&=\Gamma_{ii}^m=\left(\sum_{k=0}^{2^n-1}(-1)^{ik}p_k\right)^m=\lambda_{i}^m
\end{align}

\noindent Therefore,
\begin{equation}
    \boldsymbol{\lambda}^m=\bm{WM}^m\ket{\bm{0}}
\end{equation}

\begin{lemma} For a given $\bm{p}$, $\bm{M}$ can be written as 

\[\bm{M}=\frac{1}{2^n}\bm{W}\boldsymbol{\Gamma} \bm{W}\] 

\noindent where $\bm{W}$ is Walsh-Hadamard Transform, and $\boldsymbol{\Gamma}$ is a diagonal matrix with $ \Gamma_{ii}=\sum_{k=0}^{2^n-1}(-1)^{ik}p_k$.

\label{Lemma2}
\end{lemma}
\noindent \textit{Proof of Lemma \ref{Lemma2}:}\\
By applying Singular Value Decompostion (SVD), $\bm{M}$ can be written as $\bm{M}=\bm{V}\boldsymbol{\Gamma} \bm{V}^{-1}$ where we assume that $\bm{V}=\bm{W}$ and $\bm{V}^{-1}=\frac{1}{2^n}\bm{W}$.
The matrix $\boldsymbol{\Gamma}$ can then be written as
\begin{equation*}
\boldsymbol{\Gamma} = \bm{V}^{-1}\bm{MV}
\end{equation*}
\noindent where $
M_{ij}=p_{i\oplus j}$. We just have to prove that $\boldsymbol{\Gamma}$ is a diagonal matrix. By the definition of the Walsh-Hadamard Transform, $W_{ij}=(-1)^{ij}$,
where $ij$ is the bitwise inner product between two n-bit strings $i$ and $j$, modulo 2.
\noindent We have
\begin{align}
    \boldsymbol{\Gamma}=\frac{1}{2^n}\bm{WMW}=\frac{1}{2^n}\bm{RW}
\end{align}
where $\bm{R}=\bm{WM}$. Thus, $R_{ij}$ can be written as
\begin{equation}
R_{ij}=\sum_{r=0}^{2^{n}-1}W_{ir}M_{rj}=\sum_{r=0}^{2^{n}-1}(-1)^{ir}p_{r\oplus j}
\end{equation}

\begin{align}
    (RW)_{ij}&=\sum_{s=0}^{2^{n}-1}R_{is}W_{sj}
    =\sum_{r=0}^{2^n-1}\sum_{s=0}^{2^n-1}(-1)^{ir\oplus js}p_{r\oplus s} 
    =\sum_{k=0}^{2^n-1}(-1)^{jk}\left(\sum_{r=0}^{2^n-1}(-1)^{(i\oplus j)r}\right)p_k
\end{align}

\noindent After some simplifications,

\begin{equation}
    \Gamma_{ij}=\frac{1}{2^n}(RW)_{ij}=\left\{
\begin{array}{ll}
      \sum_{k=0}^{2^n-1}(-1)^{ik}p_k, &i=j\\
      0, & \text{otherwise}
\end{array} 
\right.
\end{equation}
Thus, $\boldsymbol{\Gamma}$ is a diagonal matrix.

\section*{Jensen-Shannon Divergence ($JSD$)}
\noindent The $JSD$ is a method used to measure the similarity between two probability distributions $\bm{p}$ and $\bm{q}$. The $JSD$ ranges between 0 and 1. The lower the $JSD$, the closer the two distributions are. The $JSD$ between $\bm{p}$ and $\bm{q}$ is calculated as
\begin{equation}
    JSD(\bm{p},\bm{q})=\frac{1}{2}D(\bm{p}||\bm{m})+\frac{1}{2}D(\bm{q}||\bm{m}) 
\end{equation}
where $ \bm{m}=\frac{1}{2}(\bm{p}+\bm{q})$ and 
\begin{equation}
    D(\bm{P},\bm{Q})=\sum_{x}\bm{P}(x) log(\frac{\bm{P}(x)}{\bm{Q}(x)})
\end{equation}

\section*{Constructing of the mitigation matrix}
The following algorithm summarizes our proposed protocol for constructing the mitigation matrix $Q_{m}$ for a new depth $m$ not included in our \textit{training set of depths}.

\begin{algorithm}
\caption{Proposed Protocol}
\begin{algorithmic}[1]

\Require Choose a \textit{training set} $\bm{T}=[1,\,m_{max}]$.

\State Choose circuit depth $m \in T$.

\State Sample a random sequence $s\in S_m$. The set of all length $m$ sequences of one-qubit Clifford gates applied independently on each qubit, followed by an inverse gate for this sequence.

\State Obtain an estimate $\hat{\bm{q}}(m,s,\ket{\bm{in}})$ of the probability distribution over the $2^n$ possible measurement outcomes.

\State Repeat steps 2-3 for $K$ times to obtain an estimate of the average probability distribution $\hat{\bm{q}}(m,\ket{\bm{in}})$ where 
\begin{equation*}
    \hat{\bm{q}}(m,\ket{\bm{in}})=\frac{1}{K}\sum{\hat{\bm{q}}}(m,s,\ket{\bm{in}})
\end{equation*}


\State Apply an input-specific permutation matrix $\boldsymbol{\Pi}_{in}$ on $\hat{\bm{q}}(m,\ket{\bm{in}})$ where
\begin{equation*}
    \Pi_{in_{ij}}=\begin{cases} 
      1 & if\,i\oplus j= in \\
      0 & otherwise
   \end{cases}
\end{equation*}

\State Apply Walsh-Hadamard transform $\bm{W}$ on $\hat{\bm{q}}(m,\ket{\bm{in}})$ to obtain 
\begin{equation*}
    \boldsymbol{\Lambda}(m)=\bm{W}\hat{\bm{q}}(m,\ket{\bm{in}})
\end{equation*}

\State Repeat steps 1-6 for all $m \in T$.

\State For each parameter $\Lambda_i(m) \in \boldsymbol{\Lambda}(m)$, fit the model
\begin{equation*}
    \Lambda_{i}(m)=A_i\lambda_{i}^m
\end{equation*}
\State Apply an Inverse Walsh-Hadamard transform on $\boldsymbol{\lambda}$ to obtain the error rates of the Pauli Channel
\begin{equation*}
    \bm{p}=\bm{W}^{-1}\boldsymbol{\lambda}
\end{equation*}

\State Construct the input-specific spam error matrix $\bm{N}_{in}=\bm{W}^{-1}\bm{AW}$ where $\bm{A}$ is a diagonal matrix with whose diagonal entries $A_{i}$. 

\State Construct the input-specific average gate error matrix $\bm{M}_{in}$ where $M_{in_{ij}}=p_{i \oplus j}$ ($\oplus$ is the bitwise exclusive-Or operator).

\State Repeat Steps 1-11 for all $\ket{\bm{in}}\in \{\ket{\bm{0}},\,\ket{\bm{1}},\,\dots,\,\ket{\bm{2^n-1}}\}$.

\State For all basis input states, estimate $\hat{\bm{q}}(m',\ket{\bm{in}})$ as
\begin{equation*}
    \bm{q}'(m',\ket{\bm{in}})=\bm{N}_{in}\bm{M}_{in}^{m'} \ket{\bm{in}}
\end{equation*}

\State Construct the mitigation matrix $\bm{Q}_{m'}$ where 
\begin{equation*}
\bm{Q}_{m'}=\left[\bm{q}'(m',\ket{\bm{0}}),\,\bm{q}'(m',\ket{\bm{1}}),\, \dots ,\,\bm{q}'(m',\ket{\bm{2^n-1}})\right]
\end{equation*}
\end{algorithmic}
\end{algorithm}

\newpage
\section*{Other Quantum Computers}
In this section, we show the results of evaluating the accuracy of the model where we include the IBM Q Athens quantum computer. The Athens quantum computer is not included in the main manuscript since it is retired by IBM.  Hence, we could not include the results of the mitigation. Similar to Figure 1 in the main manuscript, Figure 1 below  presents the computed $JSD$ for different quantum computers while varying $m_{max}$ with the addition of IBM Q Athens that demonstrates $JSD$ values that range between 0.005 and 0.08. Figure 2 presents the average and standard deviation for the test $JSD$ values for the different quantum computers including IBM Q Athens that demonstrates an average test $JSD$ between 0.02 and 0.04 for the different $m_{max}$ values. 
\begin{figure}[h]
\centering
\begin{subfigure}[h]{0.45\textwidth}
  \centering
  \includegraphics[width=\columnwidth]{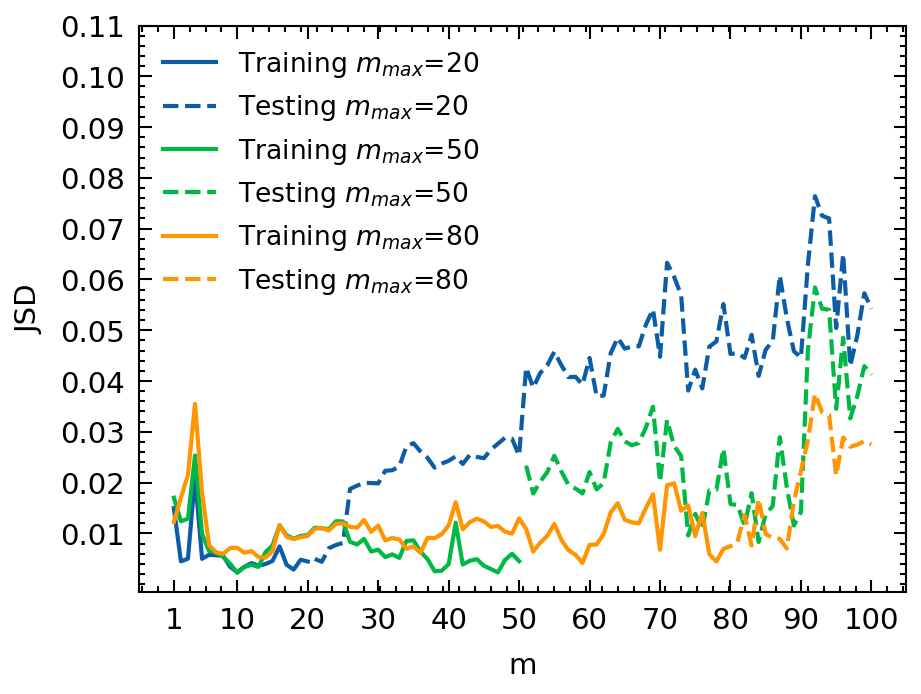}
  \caption{IBM Q Athens}
  \label{Athens}
\end{subfigure}
\begin{subfigure}[h]{0.45\textwidth}
  \centering
  \includegraphics[width=\columnwidth]{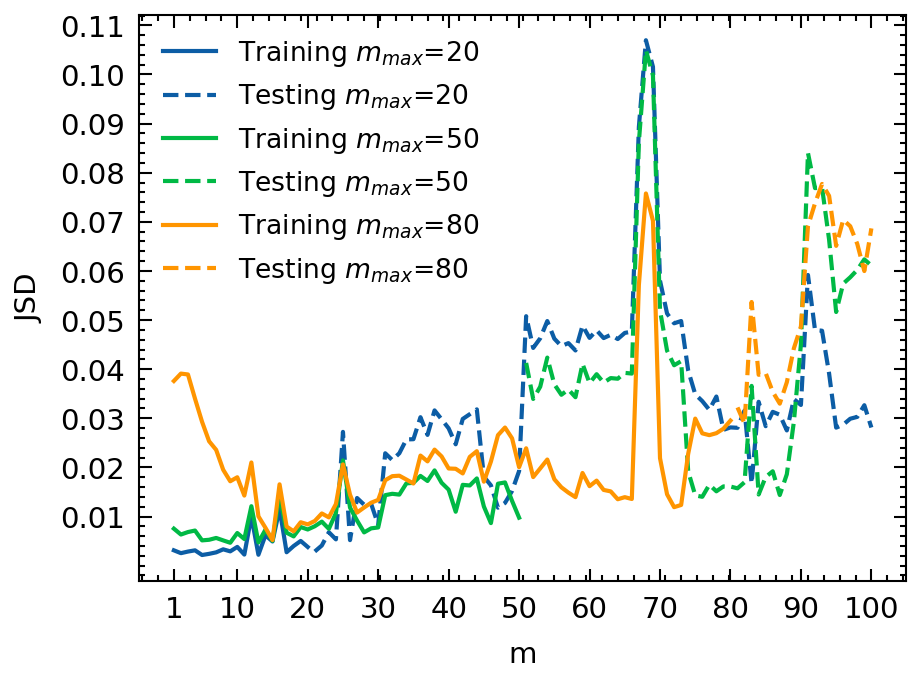}
  \caption{IBM Q Lima}
  \label{Lima}
\end{subfigure}
\begin{subfigure}[h]{0.45\textwidth}
  \centering
  \includegraphics[width=\columnwidth]{Belem}
  \caption{IBM Q Belem}
  \label{Belem}
\end{subfigure}
\begin{subfigure}[h]{0.45\textwidth}
  \centering
  \includegraphics[width=\columnwidth]{Manila}
  \caption{IBM Q Manila}
  \label{Manila}
\end{subfigure}
\caption{$JSD(\hat{\bm{q}}(m,\ket{\bm{in}}),\,\bm{q}'(m,\ket{\bm{in}}))$ for training sets of depths $T$ and testing sets of depths $T'$ with variable maximum training depth $m_{max}\in\{20,\,50,\,80\}$ on different IBMQ 5-qubit quantum computers.}
\label{VaryingTrainingDepth}
\end{figure}
\FloatBarrier
\begin{figure}[h]
    \centering
    \begin{subfigure}[h]{0.48\textwidth}
      \centering
      \includegraphics[width=\columnwidth]{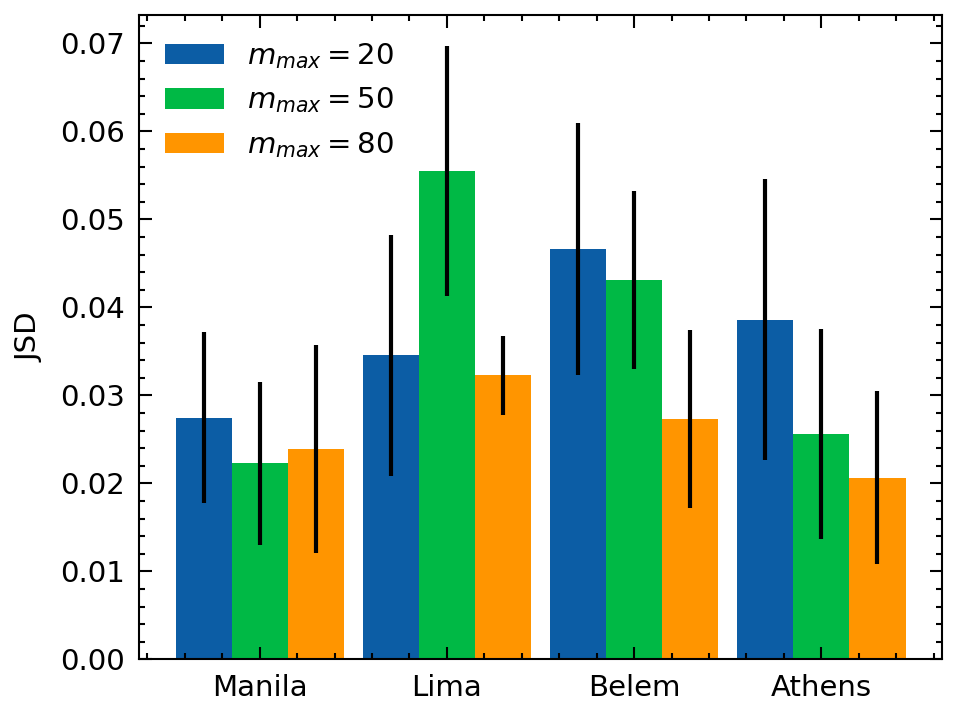}
    \end{subfigure}
    \begin{subfigure}[h]{0.48\textwidth}
      \centering
      \includegraphics[width=\columnwidth]{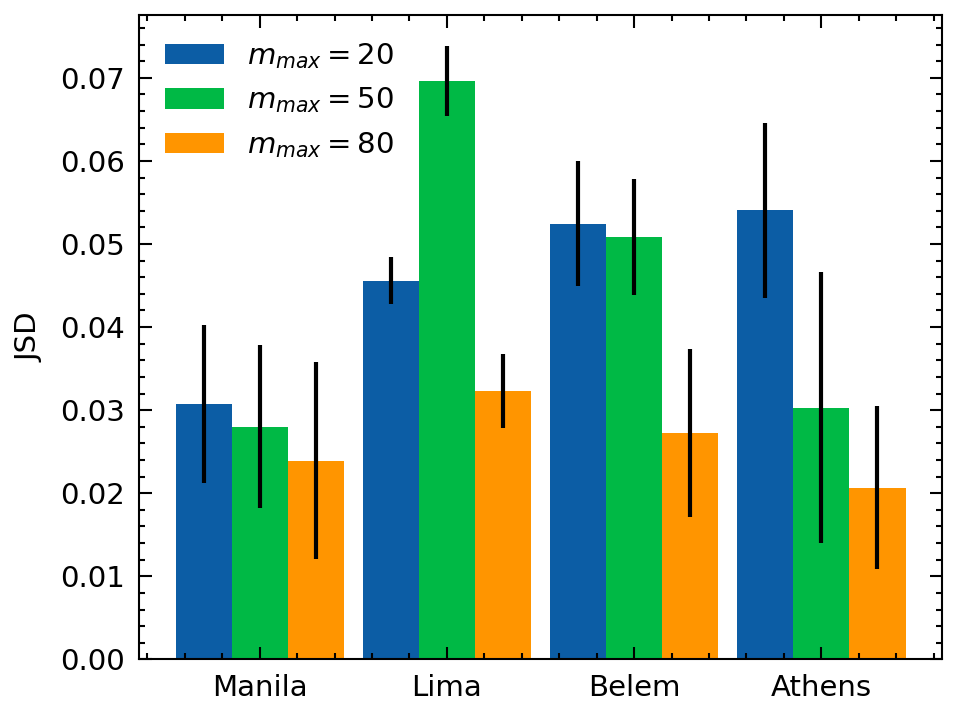}
    \end{subfigure}
    \caption{The average and standard deviation of $JSD(\hat{\bm{q}}(m,\ket{\bm{0}}),\,\bm{q}'(m,\ket{\bm{0}}))$; (a) over all depths $m \in [m_{max}+1,100]$ and  (b) over depths $m \in [80,100]$ while varying the maximum training depth $m_{max}$ on different IBM Q 5-qubit quantum computers.}
    \label{TestErrorBars}
\end{figure}

\noindent \textbf{Comparing the performance with $\bm{p}_{ave}$ vs $\bm{p}_{in}$}

\noindent In this section, we evaluate the mitigation power of $\bm{Q}_m$ when relying on the input specific error rate vectors $\bm{p}(\ket{\bm{in}})$ to construct $\bm{M}_{in}$ for all inputs compared to relying on the average error rate $\bm{p}_{avg}=\frac{1}{2^n}\sum_{in=0}^{2^n-1}{\bm{p}(\ket{\bm{in}})}$ to construct $\bm{M}_{in}=\bm{M}_{avg}$. $\bm{N}_{in}$ remains input specific. Similar to Figure 3 in the main manuscript, Figure 3 below compares the average $JSD$ between the ideal outputs and each of the unmitigated outputs, the mitigated outputs by the MEM protocol, and the mitigated outputs by $\bm{Q}_{m}$ using $\bm{p}_{ave}$ in addition to the mitigated outputs by $\bm{Q}_{m}$ using $\bm{p}(\ket{\bm{in}})$. Figure 3 shows that the difference between using $p_{ave}$ and $\bm{p}(\ket{\bm{in}})$ for the construction of $\bm{Q}_m$ is negligible for the different quantum computers indicating consistent SPAM-free error rates for all inputs.

\begin{figure}[h]
\centering
\begin{subfigure}[h]{0.48\textwidth}
  \centering
  \includegraphics[width=\columnwidth]{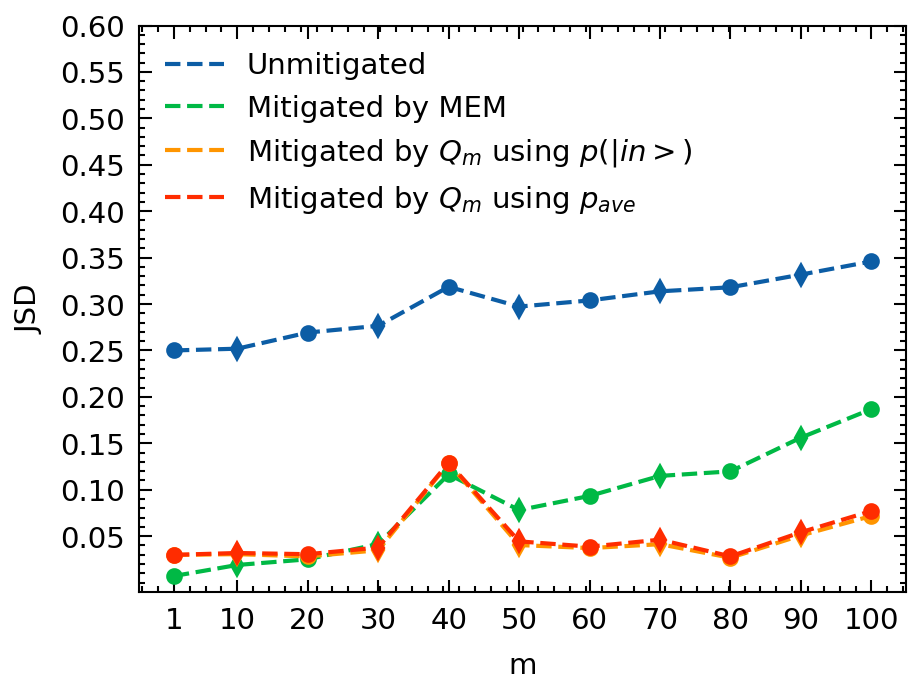}
  \caption{IBM Q Lima}
\end{subfigure}\hfill
\begin{subfigure}[h]{0.48\textwidth}
  \centering
  \includegraphics[width=\columnwidth]{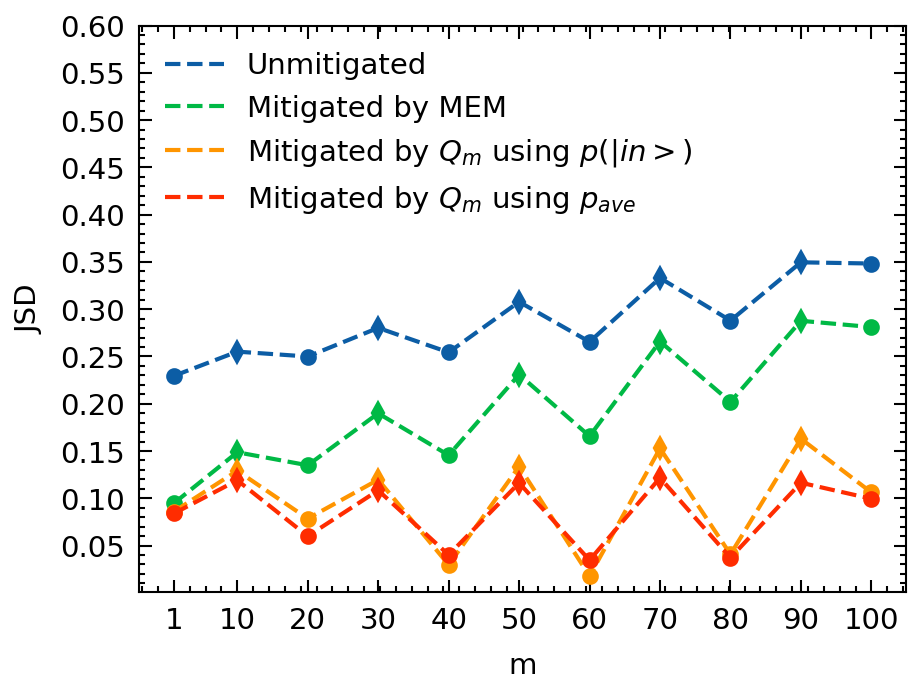}
  \caption{IBM Q Manila}
\end{subfigure}
\begin{subfigure}[h]{0.48\textwidth}
  \centering
  \includegraphics[width=\columnwidth]{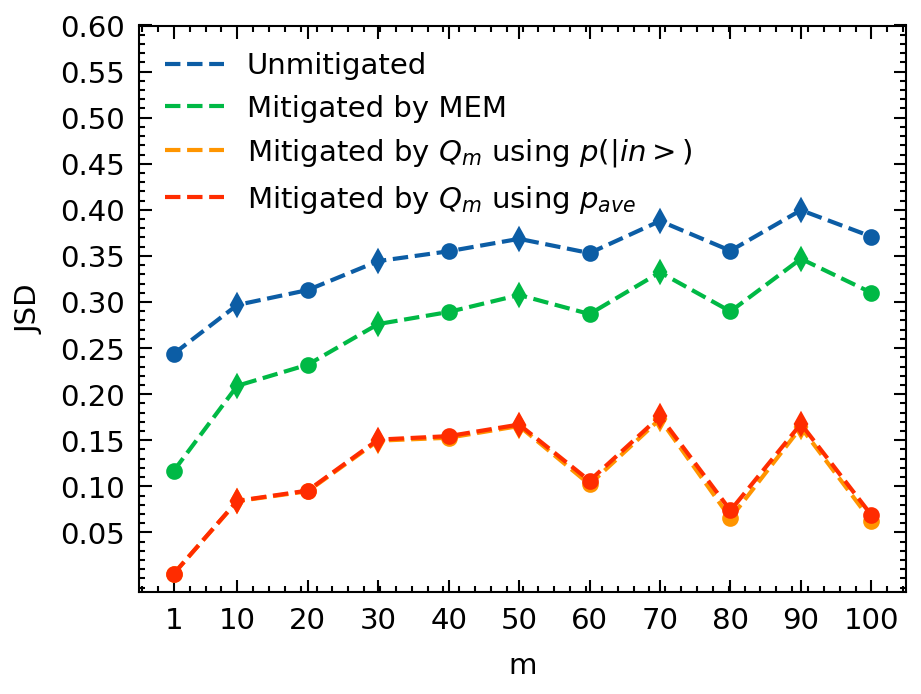}
  \caption{IBM Q Belem}
\end{subfigure}
\caption{Average $JSD$ between the ideal output $\ket{\bm{in}}$ and each of the unmitigated outputs $\hat{\bm{q}}(m,s,\ket{\bm{in}})$, mitigated outputs by the MEM protocol, and mitigated outputs by our proposed noise model using $\bm{p}(\ket{\bm{in}})$ and using $\bm{p}_{ave}$ for each depth $m$ on IBM Q 5-qubit quantum computers.}
\label{Mitigation}
\end{figure}
\FloatBarrier

\newpage
\section*{Average $JSD$ for Different Input States}
In this section, we elaborate further on Figure 3 in the main manuscript and show the average $JSD$ between the ideal output $\ket{\bm{in}}$ and the mitigated output by our proposed protocol for each depth $m$ and input state $\ket{\bm{in}}$ (Figure 4). Figure 5 presents the average and standard deviation of $JSD$ between the ideal output and mitigated output by our proposed protocol over all depths for each input state $\ket{\bm{in}}$. We notice that, for the different quantum computers, some input states demonstrate lower $JSD$ values compared to other input states. The Lima quantum computer presents the best average $JSD$ in the worst input state with a minimum of 0.0023 and a maximum of 0.107 followed by the Belem quantum computer showing a minimum of 0.074 and a maximum of 0.202 and the Manila quantum computer showing a minimum of 0.031 and a maximum of 0.280. Table 1 shows the average reduction in the test $JSD$ using the proposed protocol over the unmitigated data is about 58.33\%, 61.04\%, and 85.82\% for the Belem, Manila, and Lima computers, respectively, compared to a reduction of 16.67\%, 27.12\%, and 72.77\% for the Belem, Manila, and Lima computers, respectively, using the MEM protocol. We report upto 69\% improvement and on average 40\% improvement compared to the MEM protocol. 

\begin{figure}[h]
\centering
\begin{subfigure}[t]{0.48\textwidth}
  \centering
  \includegraphics[width=\columnwidth]{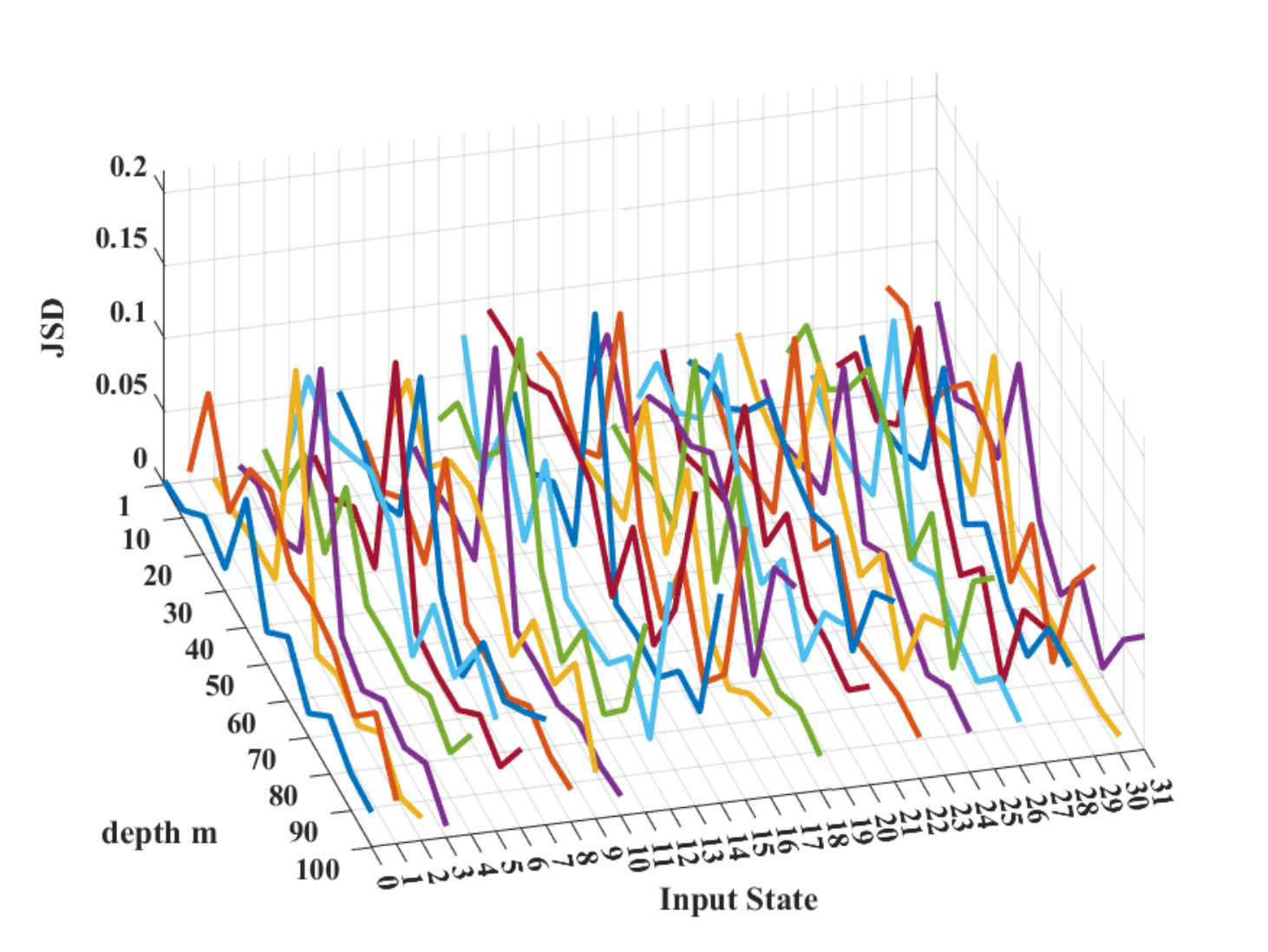}
  \caption{IBM Q Lima}
  \label{Lima} 
\end{subfigure}\hfill
\begin{subfigure}[t]{0.48\textwidth}
  \centering
  \includegraphics[width=\columnwidth]{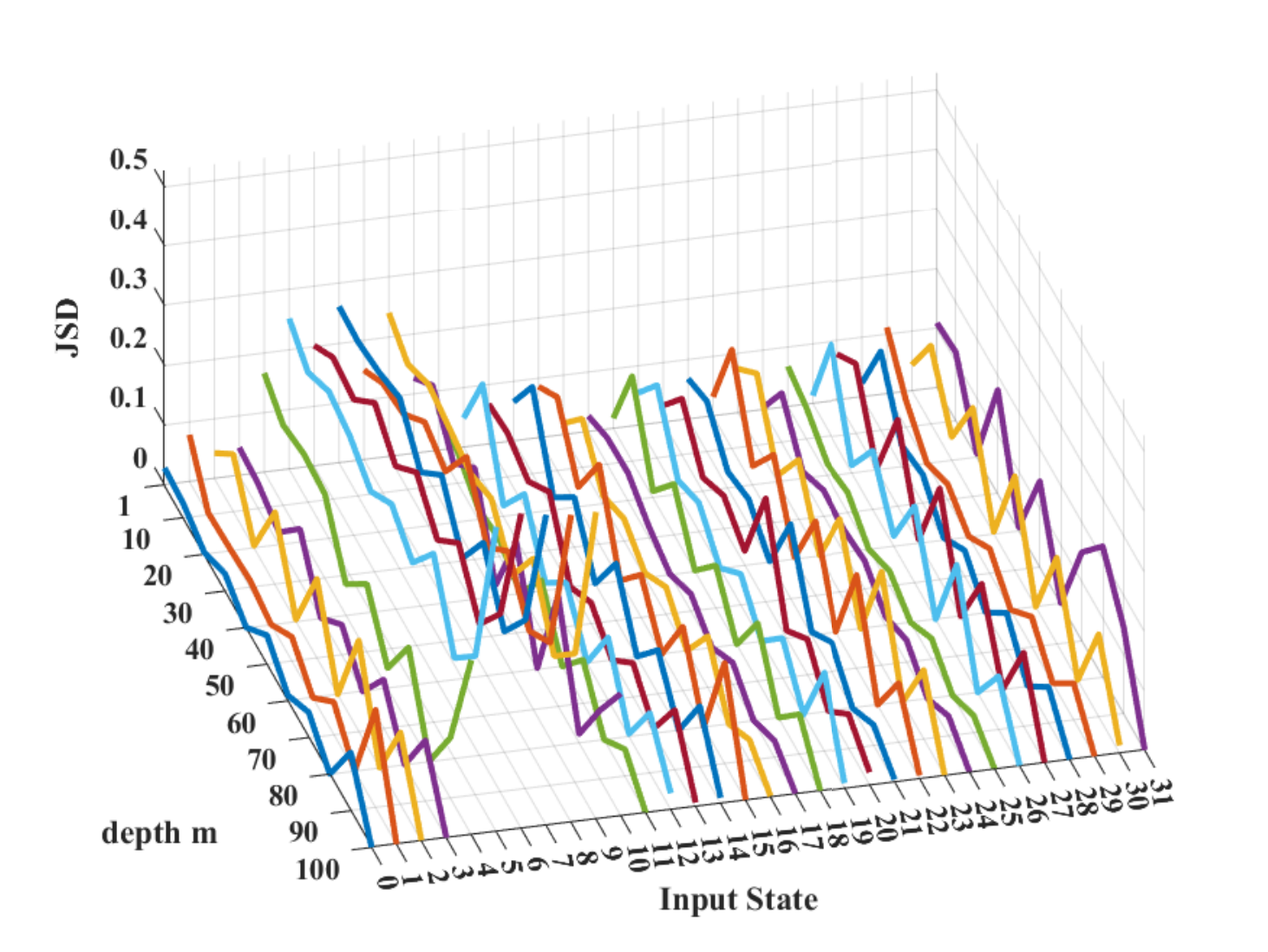}
  \caption{IBM Q Manila}
  \label{Manila}
\end{subfigure}
\begin{subfigure}[t]{0.48\textwidth}
  \centering
  \includegraphics[width=\columnwidth]{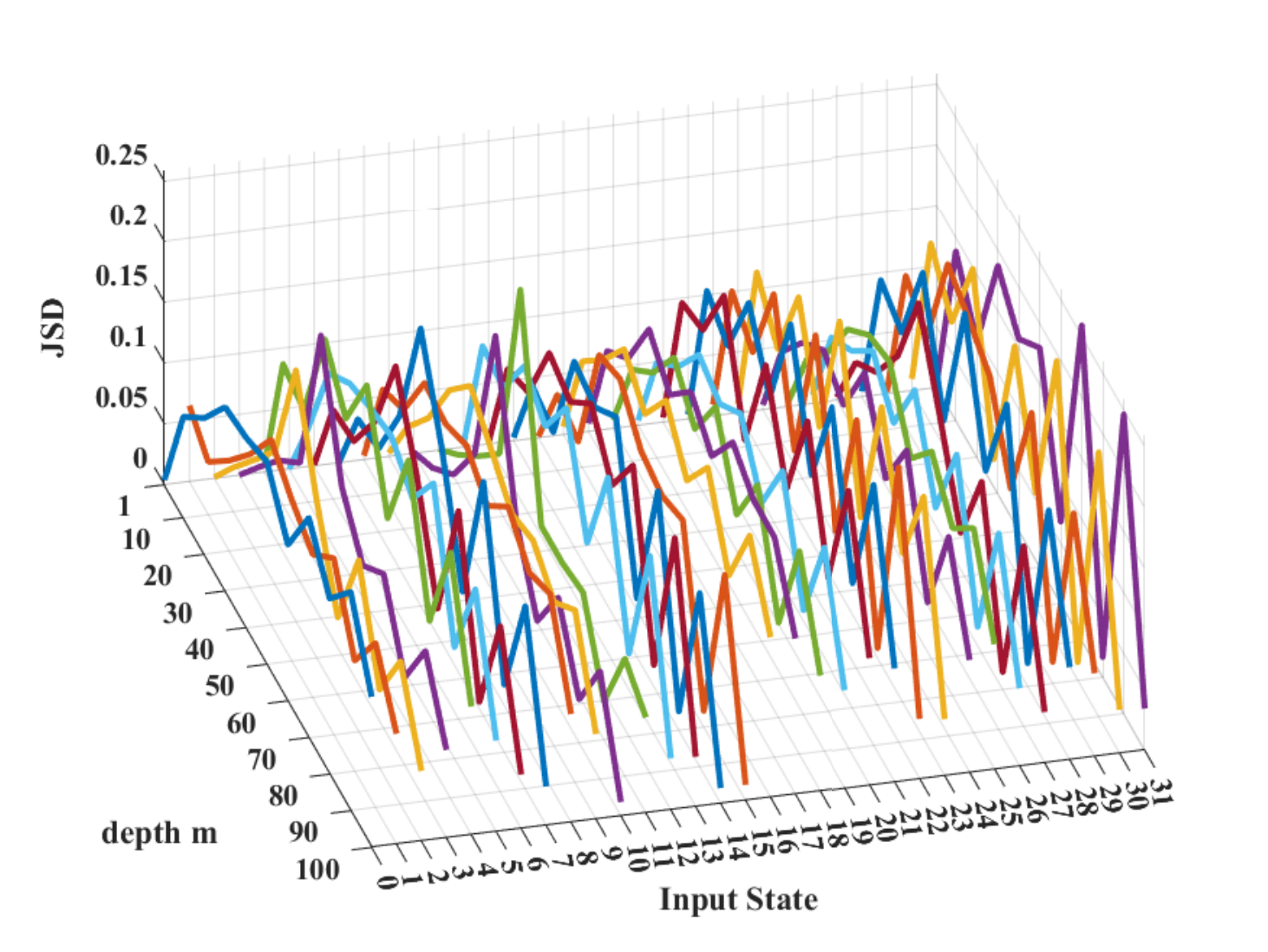}
  \caption{IBM Q Belem}
  \label{Belem}
\end{subfigure}
\caption{Average $JSD$ between the ideal output $\ket{\bm{in}}$ and the mitigated outputs by our proposed noise model for each depth $m$ and each input state $\ket{\bm{in}}$ on IBM Q 5-qubit quantum computers.}
\label{VaryingTrainingDepth}
\end{figure}
\FloatBarrier

\begin{figure}[h]
\centering
\begin{subfigure}[t]{0.5\textwidth}
  \centering
  \includegraphics[width=\columnwidth]{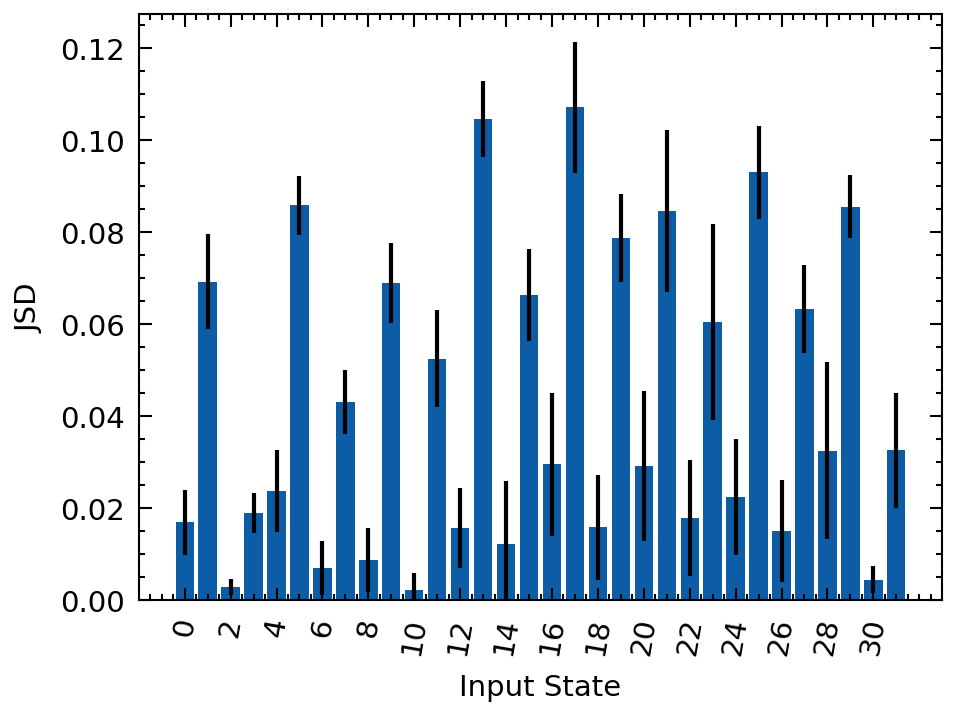}
  \caption{IBM Q Lima}
  \label{Lima} 
\end{subfigure}\hfill
\begin{subfigure}[t]{0.5\textwidth}
  \centering
  \includegraphics[width=\columnwidth]{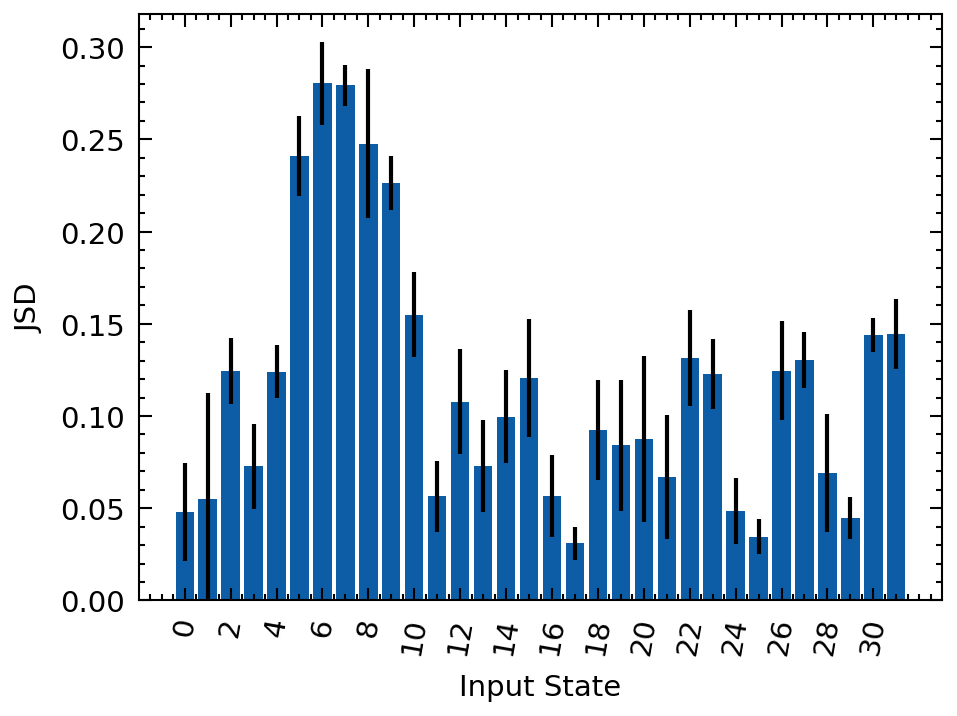}
  \caption{IBM Q Manila}
  \label{Manila}
\end{subfigure}
\begin{subfigure}[t]{0.5\textwidth}
  \centering
  \includegraphics[width=\columnwidth]{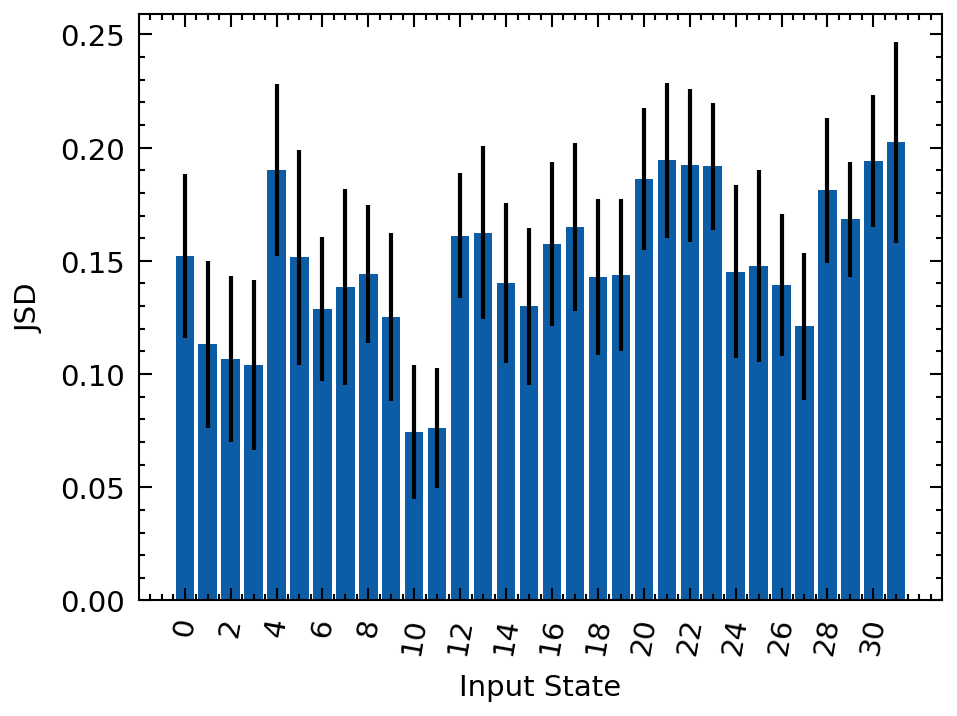}
  \caption{IBM Q Belem}
  \label{Belem}
\end{subfigure}
\caption{Average and standard deviation of $JSD$ over all depths $m$ between the ideal output $\ket{\bm{in}}$ and the mitigated output by our proposed noise model for each input state $\ket{\bm{in}}$ on IBM Q 5-qubit quantum computers.}
\label{VaryingTrainingDepth}
\end{figure}
\FloatBarrier

\begin{table}[]
\centering
\caption{Average test $JSD$ between the ideal output $\ket{\bm{in}}$ and each of the unmitigated output, mitigated by the proposed protocol output, and mitigated by the MEM protocol output for the different quantum computers.}
\label{tab:my-table}
\resizebox{\textwidth}{!}{%
\begin{tabular}{|c|c|c|c|c|c|c|}
\hline
&m & Unmitigated JSD & MEM JSD & Mitigated by Proposed Protcol JSD & MEM improvement \% & Proposed improvement \% \\ \hline \hline
\parbox[t]{2mm}{\multirow{7}{*}{\rotatebox[origin=c]{90}{Belem}}} &10 & 0.30 & 0.21 & 0.08 & 30.00 & 73.33 \\ \cline{2-7}
&30 & 0.34 & 0.28 & 0.15 & 17.65 & 55.88 \\ \cline{2-7}
&50 & 0.37 & 0.31 & 0.17 & 16.22 & 54.05 \\ \cline{2-7}
&70 & 0.39 & 0.33 & 0.18 & 15.38 & 53.85 \\ \cline{2-7}
&90 & 0.40 & 0.35 & 0.17 & 12.50 & 57.50 \\ \cline{2-7}
&mean & 0.36 & 0.30 & 0.15 & 16.67 & 58.33 \\ \hline
\hline
\parbox[t]{2mm}{\multirow{7}{*}{\rotatebox[origin=c]{90}{Manila}}} &10 & 0.26 & 0.15 & 0.12 & 42.31 & 53.85 \\ \cline{2-7}
&30 & 0.28 & 0.19 & 0.11 & 32.14 & 60.71 \\ \cline{2-7}
&50 & 0.31 & 0.23 & 0.12 & 25.81 & 61.29 \\ \cline{2-7}
&70 & 0.33 & 0.27 & 0.12 & 18.18 & 63.64 \\ \cline{2-7}
&90 & 0.35 & 0.29 & 0.12 & 17.14 & 65.71 \\ \cline{2-7}
&mean & 0.31 & 0.23 & 0.12 & 27.12 & 61.04 \\ \hline \hline
\parbox[t]{2mm}{\multirow{7}{*}{\rotatebox[origin=c]{90}{Lima}}} &10 & 0.25 & 0.02 & 0.03 & 92.00 & 88.00 \\ \cline{2-7}
&30 & 0.28 & 0.04 & 0.04 & 85.71 & 85.71 \\ \cline{2-7}
&50 & 0.30 & 0.08 & 0.04 & 73.33 & 86.67 \\ \cline{2-7}
&70 & 0.31 & 0.12 & 0.05 & 61.29 & 83.87 \\ \cline{2-7}
&90 & 0.33 & 0.16 & 0.05 & 51.52 & 84.85 \\ \cline{2-7}
&mean & 0.29 & 0.08 & 0.04 & 72.77 & 85.82 \\ \hline
\end{tabular}%
}
\end{table}


\end{document}